\documentclass{webofc}
\usepackage[varg]{txfonts}   % Web of Conferences font

\usepackage{times}
\usepackage{xspace} % important for correct spacing - see command list

\usepackage{chngcntr}
\usepackage{graphicx} 
\usepackage{bm}
\usepackage{amssymb}  
\usepackage{slashed}
\usepackage{booktabs}
\usepackage{multirow}
\usepackage{mathtools}
\usepackage{amsmath}
\usepackage{stackrel}
\usepackage{rotating}
\usepackage{rotfloat}
\usepackage{hyperref}
\hypersetup{
    colorlinks=true,       % false: boxed links; true: colored links
    linkcolor=blue,          % color of internal links
    citecolor=blue,        % color of links to bibliography
    filecolor=blue,      % color of file links
    urlcolor=blue           % color of external links
}
\usepackage{amsmath}
\usepackage[utf8]{inputenc}
\usepackage{cleveref}
\usepackage{makecell}
\usepackage{cellspace} %
\allowdisplaybreaks

\newcommand{\ga}{\gamma}

\newcommand{\bi}{\begin{itemize}}
\newcommand{\ei}{\end{itemize}}

\newcommand{\ben}{\begin{enumerate}}
\newcommand{\een}{\end{enumerate}} 

\newcommand{\bt}[1]{\begin{table}[tb]\begin{tabular}{#1} \hline\hline  \\[-1.0em]}
\newcommand{\et}[2]{\hline\hline \end{tabular} \caption{#1} \label{#2} \end{table}}

\newcommand{\be}{\begin{equation}}
\newcommand{\ee}{\end{equation}}

\newcommand{\bea}{\begin{eqnarray}}
\newcommand{\eea}{\end{eqnarray}}

\newcommand{\gaga}{\gamma\gamma}
\newcommand{\gagagaga}{\ensuremath{\gamma\gamma\rightarrow\gamma\gamma}\xspace}

\newcommand{\mev}{\ensuremath{\mathrm{\,Me\kern -0.1em V}}\xspace}
\newcommand{\gev}{\ensuremath{\mathrm{\,Ge\kern -0.1em V}}\xspace}

\begin{document}
\title{Two-photon decay of fully-charmed tetraquarks from  light-by-light scattering at the LHC}

\author{
    \firstname{Volodymyr} \lastname{Biloshytskyi}
    \inst{1}
    \fnsep
    \thanks{Speaker, \email{vbiloshy@uni-mainz.de}} 
\and
    \firstname{Lucian} \lastname{Harland-Lang}
    \inst{2}
    \fnsep
\and
    \firstname{Bogdan} \lastname{Malaescu}
    \inst{3}
    \fnsep
\and
    \firstname{Vladimir} \lastname{Pascalutsa}
    \inst{1}
    \fnsep
\and
    \firstname{Kristof} \lastname{Schmieden}
    \inst{4}
    \fnsep
\and
    \firstname{Matthias} \lastname{Schott}
    \inst{4}
    \fnsep
}

\institute{Institut f\"ur Kernphysik,
 Johannes Gutenberg-Universit\"at  Mainz,  D-55128 Mainz, Germany
\and
    Rudolf Peierls Centre, Beecroft Building, Parks Road, Oxford, OX1 3PU, United Kingdom
\and
   LPNHE, Sorbonne Universit\'e, Universit\'e Paris Cit\'e, CNRS/IN2P3, Paris, France, 75252
\and
Institut f\"ur Physik,
 Johannes Gutenberg-Universit\"at  Mainz,  D-55128 Mainz, Germany
          }

\abstract{
The LHC newly-discovered resonant structures around 7 GeV, such as the $X(6900)$,
could be responsible for the observed excess in light-by-light scattering
between 5 and 10 GeV. 
We show that the ATLAS data for light-by-light scattering may indeed be explained
by such a state with the $\gamma \gamma$ branching ratio of order of $10^{-4}$. This is much larger than the value inferred by the vector-meson dominance, but agrees quite well with the tetraquark expectation for the nature of this state. Further light-by-light scattering data in this region, obtained during the ongoing Run-3 and future Run-4 of the LHC,  are required to pin down these states in $\gamma\gamma$ channel.
}
\maketitle

In 2020 the ATLAS Collaboration provided the most comprehensive dataset of the observation of light-by-light (LbL) scattering in the ultra-peripheral Pb-Pb collisions from the LHC Run-2 \cite{ATLAS2020}. The statistics has been increased compared to the first analyses \cite{ATLAS2017,CMS2019}, and the unfolded data were provided. The new results show a mild excess over the Standard Model prediction centered on the diphoton invariant mass region of 5 to 10 GeV. In terms of the total LbL cross section, the discrepancy between the experimental measurement and the theoretical estimation reaches the value of around $2\sigma$.

On the other hand, the LHCb Collaboration discovered  a new state, $X(6900)$,  seen in the di-$J/\psi$ spectrum at around 6.9 GeV. This state has been confirmed very recently at ATLAS \cite{ATLAS:2022hhx} and CMS \cite{CMS:2022yhl}.
Apart from $X(6900)$, these collaborations observed two other resonances in the vicinity, namely $X(6600)$ and $X(7300)$. They all are candidates for 
fully-charmed tetraquarks, predicted in many quark models, see, e.g. \cite{Richard2020,Sonnenschein2020,Faustov2021,Deng2020,Guo2020,Barnea2006,Berezhnoy2011,Karliner2016,Wang2017,Liu2019,Weng2020,Lundhammar2020,Wan2020,Zhu2020,Liang2021,Wang:2020wrp,Wang:2022jmb,Dong:2020nwy,Chen2022}. 

The quantum-number assignment for these stated is yet to be done; the most likely options are:  scalars ($J^{PC}=0^{++}$), pseudoscalars ($J^{PC}=0^{-+}$), axial vectors ($J^{PC}=1^{-+}$), and tensors ($J^{PC}=2^{\pm+}$). 
In cases of even spin, the new $X$-resonances would couple to two photons and hence contribute to the LbL scattering. From this point of view, the $\gagagaga$ channel can be used as a filter for the resonances with even spin. Such resonances can in principle be responsible for the aforementioned discrepancy between theory and experiment observed in LbL scattering by the ATLAS Collaboration.  

\begin{figure}[tbh]
    \centering
    \includegraphics[width=0.5\linewidth]{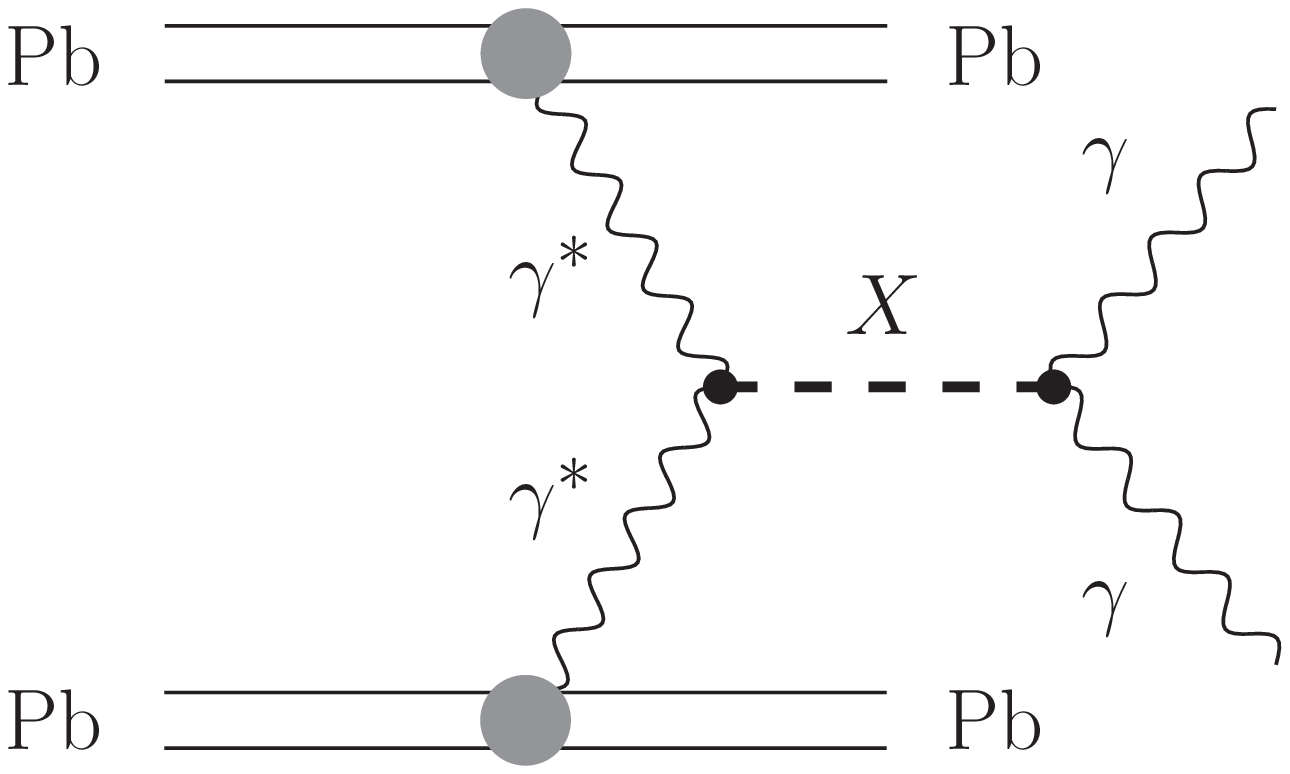}
   \hskip2cm \includegraphics[width=0.3\linewidth]{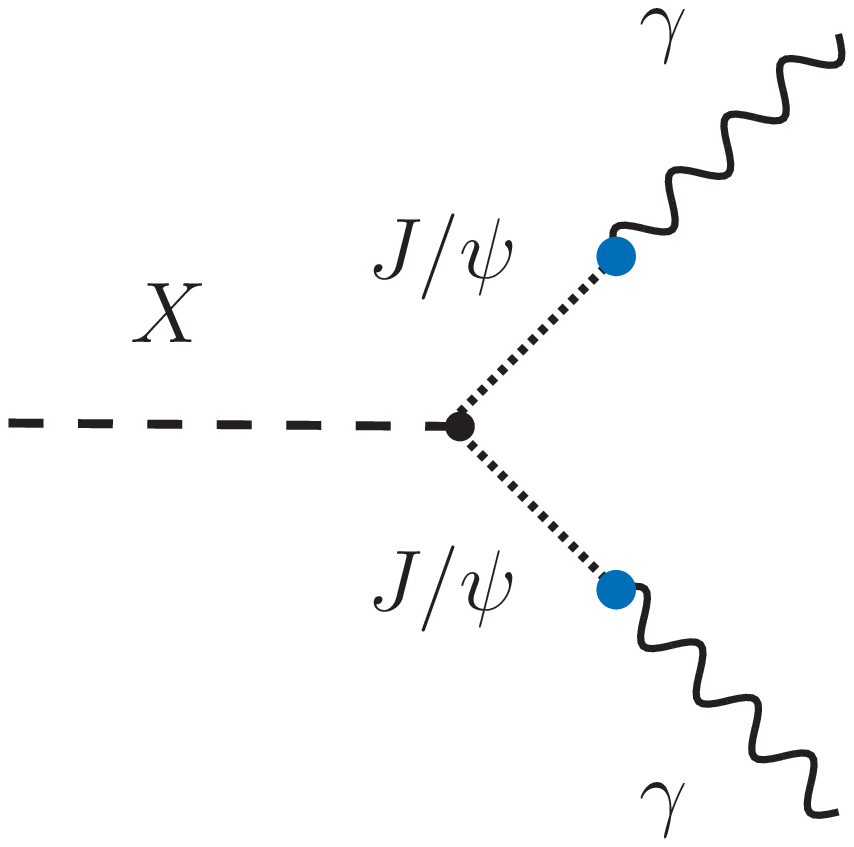}
    \caption{Left: the $X$-resonance contribution to light-by-light scattering in ultraperipheral heavy-ion collisions (crossed graphs are omitted). Right:
    the vector-meson-dominance (VMD) mechanism for the two-photon decay.}
    \label{fig:UPCyy}
\end{figure}

We have considered this scenario in the recent paper \cite{Biloshytskyi:2022dmo}, by including the $X(6900)$ contribution to the LbL scattering, as shown in Fig~\ref{fig:UPCyy}.
To this end, we extended the Monte-Carlo (MC) code SuperChic v3.05 \cite{HarlandLang2019,HarlandLang2016}\footnote{The most recent version concerning the LbL channel.}
used in the original interpretation of the ATLAS data \cite{ATLAS2020}, by including the $X(6900)$ along with the well-known bottomonium states \cite{Wang2018} pertinent to this energy region, i.e. $\eta_b$(1S), $\eta_b$(2S), $\chi_{b0}$(1P) and $\chi_{b0}$(2P).
% see Table~\ref{tab:bottom_include}.    
Note that SuperChic v3.05 includes otherwise 
only the simplest perturbative-QCD  contributions to LbL scattering, i.e., the quark-loop contribution.  The next-to-leading order corrections were shown to contribute at the order of few percent \cite{Bern:2001dg,Klusek-Gawenda:2016nuo,Krintiras:2022jxa}, which is negligible at the current level of experimental precision.

\begin{figure*}
    \centering
  \includegraphics[width=0.48\linewidth]{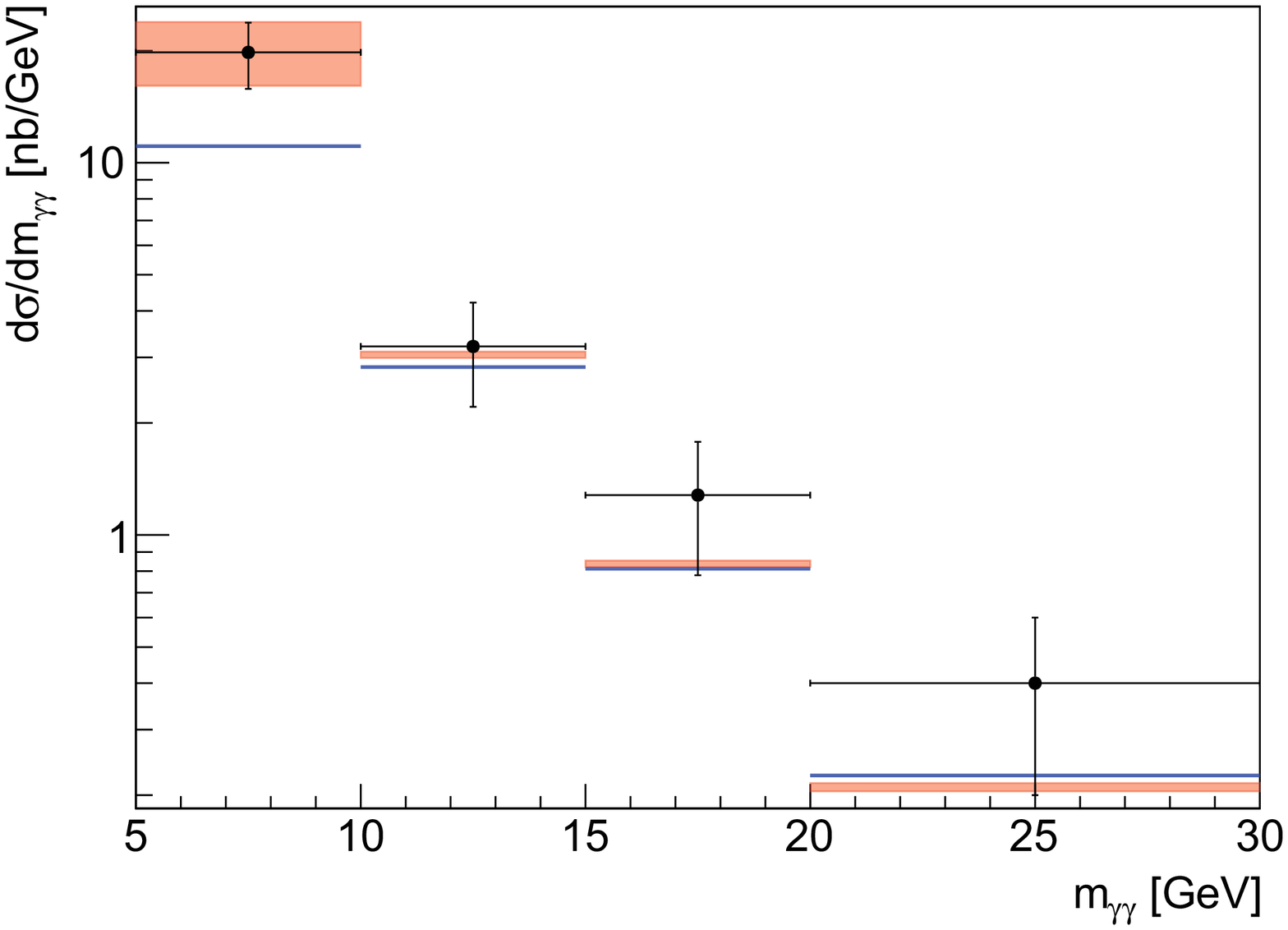}
  \includegraphics[width=0.48\linewidth]{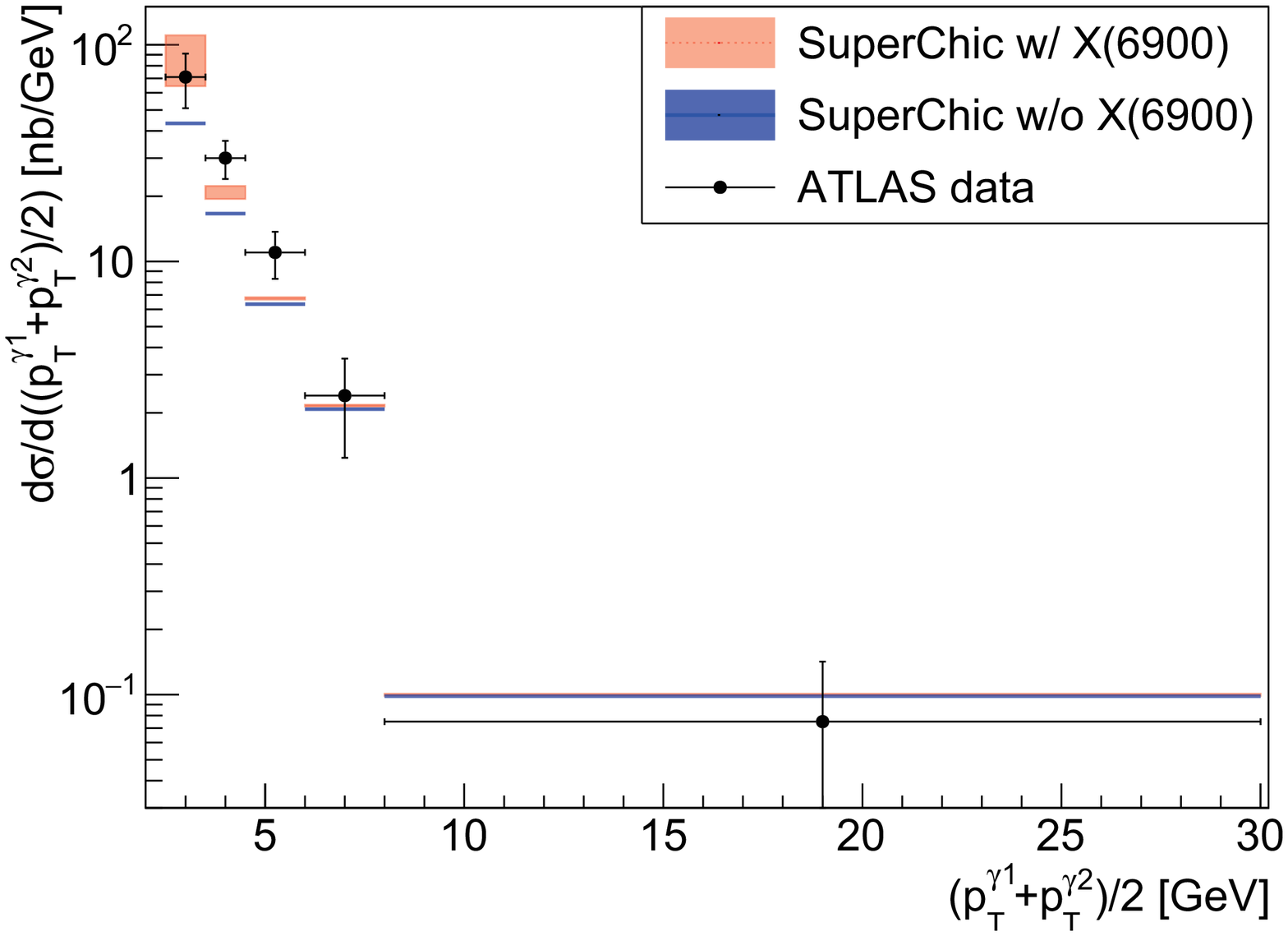}
  \includegraphics[width=0.48\linewidth]{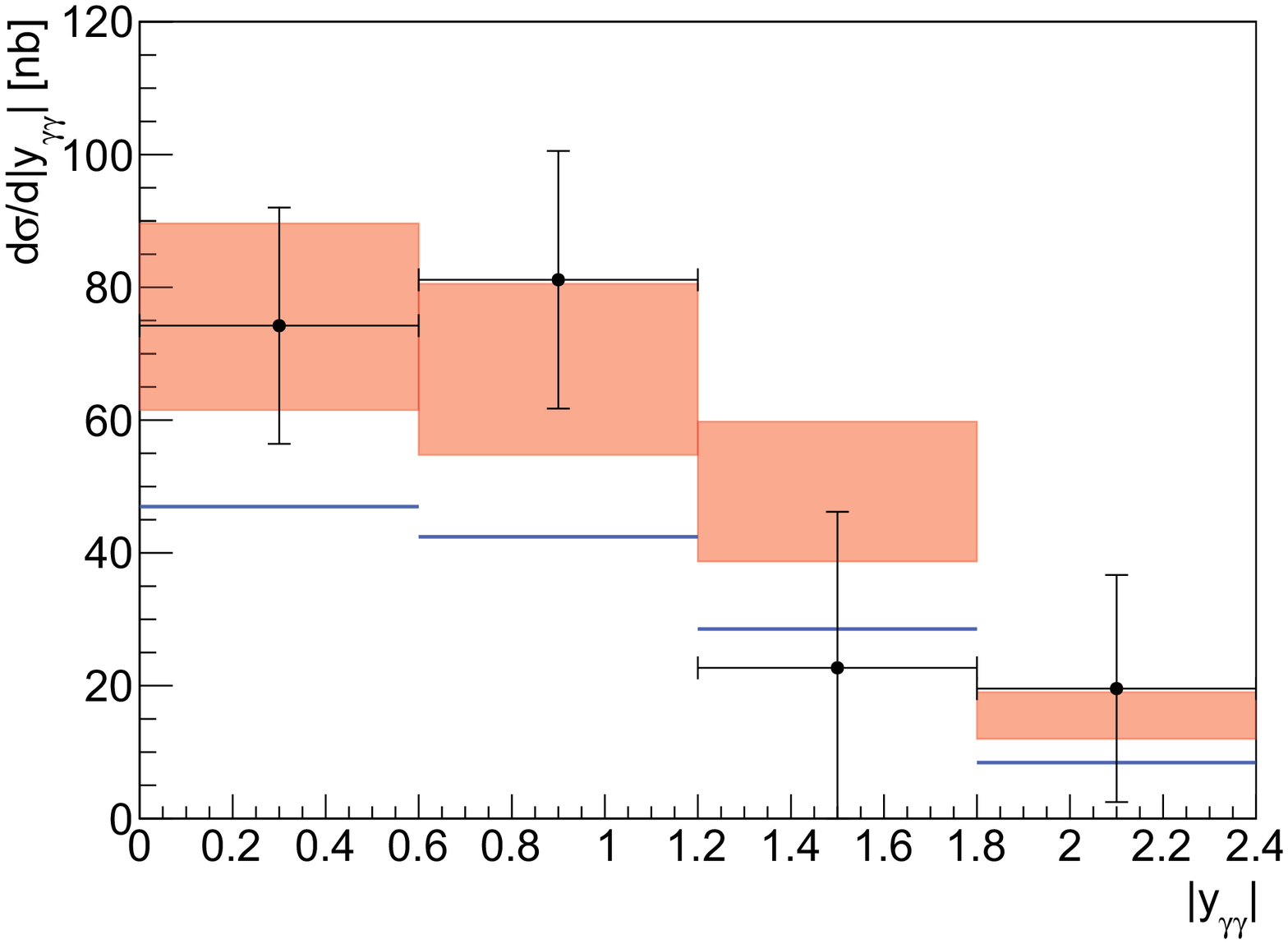}
  \includegraphics[width=0.48\linewidth]{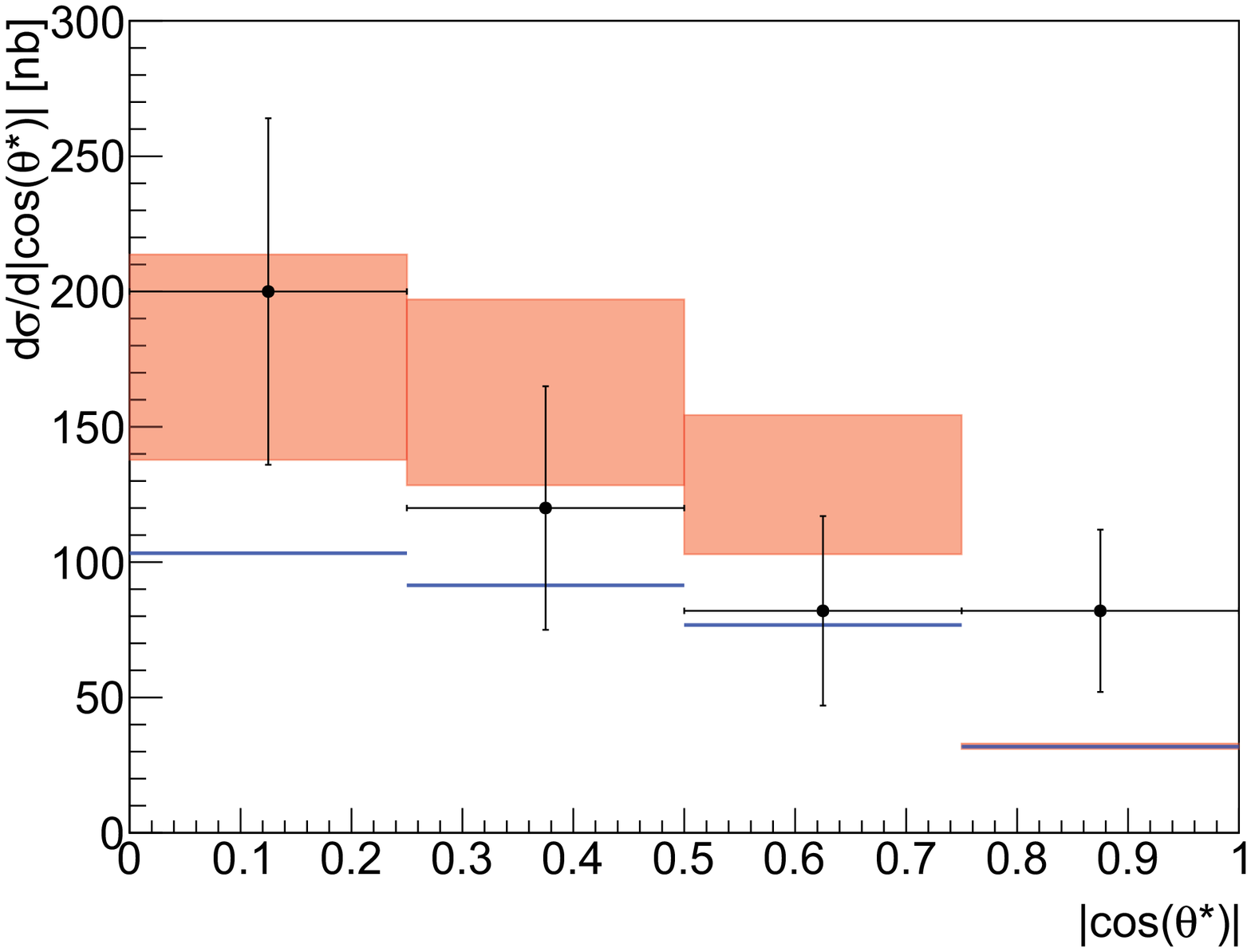}
    \caption{Differential fiducial cross sections of $\gagagaga$ production in Pb$+$Pb collisions at a centre-of-mass energy of $\sqrt{s_{NN}}=5.02$ TeV with integrated luminositiy 2.2 nb$^{-1}$ for four observables (from left to right and top to bottom): diphoton invariant mass $m_{\gaga}$, diphoton absolute rapidity $|y_{\gaga}|$, average photon transverse momentum $(p_{T}^{\ga_1}+p_{T}^{\ga_2})/2$ and diphoton $|cos(\theta^{\ast})|\equiv|\tanh(\Delta y_{\ga_1,\ga_2}/2)|$. The red band represents an uncertainty ($1\sigma$ range) of the fit with $X(6900)$. The blue band contains only the statistical uncertainty of the SuperChic simulation without $X$-resonance.}
    \label{fig:MainFigure}
\end{figure*}

The MC generator SuperChic, extended by $X(6900)$, has been used to fit
 the resonance parameters into the ATLAS LbL data. Given the mass and width of $X(6900)$ from the LHCb determination, we have determined the two-photon-decay width $\Gamma_{X\to\gaga}$, with the assumption that the total width is  dominated by the di-$J/\psi$ decay (i.e., $\Gamma_\mathrm{tot}\simeq \Gamma_{X\to J/\psi\, J/\psi}$).
The fit has been performed to the unfolded diphoton invariant mass spectrum of the ATLAS data. The CMS data is not used  in the present analysis since the corresponding spectrum is not unfolded.
We have explored both the scalar and pseudoscalar nature of $X(6900)$, but the corresponding results of the fit turn out to be  indistinguishable at the current level of statistical accuracy. 
The results for the two-photon width $\Gamma_{X\to\gamma\gamma}$ and the corresponding branching ratio are given in Table~\ref{tab:fitresults}, for  the two scenarios (interference, no-interference) considered in Ref.~\cite{LHCb2020}.
\begin{table}[h]
\centering\setcellgapes{4pt}\makegapedcells
\caption{\label{tab:fitresults} The two-photon width and the corresponding branching ratio of X(6900) obtained in \cite{Biloshytskyi:2022dmo} by fitting the light-by-light scattering data of Ref.~\cite{ATLAS2020}.}
\begin{tabular}{c c c}
\hline
    Parameter & Interference & No-interference\\
    \hline
    $\Gamma_{X(6900)\to \gamma\gamma}$ [keV]  & $67^{+15}_{-19} $ & $45^{+11}_{-14} $ \\
    $B[X(6900)\to \gamma\gamma]$  & $ 4.0^{+0.9}_{-1.1} \times10^{-4}$ & $5.6^{+1.3}_{-1.6}\times10^{-4}$ \\
    \hline
\end{tabular}
 \end{table} 
 The corresponding differential observables are shown on Fig.~\ref{fig:MainFigure}. The plots demonstrate that the inclusion of the resonance improves the description of each of the observables.

The extracted width in Table~\ref{tab:fitresults} can be compared to estimates
based on the vector meson dominance (VMD) \cite{Barger1975,Redlich2000} (see also \cite{Biloshytskyi:2022dmo} for more specifics), given in Table~\ref{tab:VMD_CMS}, for all three
newly discovered states, assuming they are scalar or pseudoscalar.
One can see that VMD predicts a much smaller branching ratio for $X(6900)$ than required to remedy the ATLAS discrepancy in the LbL cahnnel; although, given the large uncertainties, the difference is not yet very significant statistically. 

\begin{table}[H]
\centering\setcellgapes{4pt}\makegapedcells
\caption{\label{tab:VMD_CMS} Our VMD estimate of two-photon branching ratios for resonances detected at CMS\cite{CMS:2022yhl}.}
\begin{tabular}{ c  c  c  c  }
\hline
    & $B_\mathrm{VMD}(X(6600)\to\gamma\gamma)$ & $B_\mathrm{VMD}(X(6900)\to\gamma\gamma)$ & $B_\mathrm{VMD}(X(7300)\to\gamma\gamma)$\\
    \hline
    scalar & $(4.12\pm0.54)\times10^{-6}$ & $(2.77\pm0.36)\times10^{-6}$ & $(2.19\pm0.28)\times10^{-6}$ \\
   pseudoscalar &  $(15.7\pm2.2)\times10^{-6}$ & $(6.07\pm0.80)\times10^{-6}$ & $(3.73\pm0.49)\times10^{-6}$ \\
    \hline
\end{tabular}
\end{table}
%However, the contribution of $X(6600)$ and $X(7300)$ in addition to $X(6900)$ with the VMD-estimated two-photon decay widths into the $\gagagaga$ channel is still indistinguishable from the LbL background.

An alternative estimate is provided in the quark-model picture (see, e.g., \cite{Esposito2021}), assuming that the tetraquark  state forms an unstable $cc$-$\bar{c}\bar{c}$ diquark-antidiquark configuration. For the state of two charged scalar equall-mass consituents,  we apply the non-relativistic approximate formula for the two-photon decay,
\begin{equation}
    \Gamma_{X\to\gaga} = \left(\frac{4}{3}\right)^4\frac{4\pi\alpha^2}{m_{X}^2}|\psi(0)|^2,
\end{equation}
where the factor 4/3 stems from the diquark charge in units of $e$.
The value of the bound-state wave function at the origin, $\psi(0)$, can be taken from a quark model (e.g.,  Table~3 of \cite{Debastiani2019}: $|\psi(0)|^2 \simeq 2.85/4\pi$ GeV$^3$). This estimate yields the two-photon width of order of 10 keV, which is substantially larger than VMD and already much closer to the values extracted from the fit to the LbL data.

Certainly, the ongoing Run-3 measurements of both the di-$J/\psi$ and $\gamma \gamma$ channels, with increased resolution, will pin down the apparent conflict between the VMD and quark-model expectations. The possible underestimation of the VMD compared to the fitting values can be explained by the peculiar internal structure of the states, or, alternatively, by other exotic resonances in the diphoton mass region from 5 to 10 GeV, which contribute to the observed excess in the $\gamma \gamma$ channel. 

In conclusion, we have shown that the new tetraquark state $X(6900)$, observed by LHCb Collaboartion in the di-$J/\psi$ channel, as well as the other new states,  observed recently by CMS and ATLAS Collaborations,
could, in principle, account for the observed excess in the light-by-light scattering between 5 and 10 GeV. 
The inclusion of $X(6900)$ brings the Standard Model prediction in agreement with LbL data by fitting  the $X(6900)\to \gamma\gamma$ decay width.
The result presented in Table~\ref{tab:fitresults} exceeds the VMD estimates,
but is roughly in agreement with quark-model expectation of the fully-charmed tetraquark. Further measurements of the LbL scattering in the 5 to 10 GeV diphoton-mass range with increased resolution are very desirable
to improve the precision and resolve these new states. The prospective double-differential (or even triple-differential) measurements of a pair (or triplet) of the observables depicted in Fig.~\ref{fig:MainFigure}
% , which show complementary sensitivity to $X$-state, 
may provide additional information.
Future measurements that allow for the partial-wave analysis and the quantum-number assignment should be pursued by all of the LHC collaborations.

\section*{Acknowledgments}

This work was supported by the Deutsche Forschungsgemeinschaft (DFG) through the Research Unit FOR5327 [Photon-photon interactions in the Standard Model and beyond, Projektnummer 458854507]. L.H.L. thanks the Science and Technology Facilities Council (STFC) for support via grant awards ST/L000377/1 and ST/T000864/1.

\bibliography{LbL_LHC}

\begin{thebibliography}{36}

\bibitem{ATLAS2020}
G.~Aad et~al. (ATLAS), JHEP \textbf{03}, 243 (2021)

\bibitem{ATLAS2017}
M.~Aaboud et~al. (ATLAS), Nature Phys. \textbf{13}, 852 (2017)

\bibitem{CMS2019}
A.M. Sirunyan et~al. (CMS), Phys. Lett. B \textbf{797}, 134826 (2019)

\bibitem{ATLAS:2022hhx}
{ATLAS collaboration} (2022), {ATLAS-CONF-2022-040},
  \urlstyle{tt}\url{https://cds.cern.ch/record/2815676}

\bibitem{CMS:2022yhl}
{CMS collaboration} (2022), {CMS-PAS-BPH-21-003},
  \urlstyle{tt}\url{https://cds.cern.ch/record/2815336}

\bibitem{Richard2020}
J.M. Richard, Sci. Bull. \textbf{65}, 1954 (2020)

\bibitem{Sonnenschein2020}
J.~Sonnenschein, D.~Weissman, Eur. Phys. J. C \textbf{81}, 25 (2021)

\bibitem{Faustov2021}
R.N. Faustov, V.O. Galkin, E.M. Savchenko, Universe \textbf{7}, 94 (2021)

\bibitem{Deng2020}
C.~Deng, H.~Chen, J.~Ping, Phys. Rev. D \textbf{103}, 014001 (2021)

\bibitem{Guo2020}
Z.H. Guo, J.A. Oller, Phys. Rev. D \textbf{103}, 034024 (2021)

\bibitem{Barnea2006}
N.~Barnea, J.~Vijande, A.~Valcarce, Phys. Rev. D \textbf{73}, 054004 (2006)

\bibitem{Berezhnoy2011}
A.V. Berezhnoy, A.V. Luchinsky, A.A. Novoselov, Phys. Rev. D \textbf{86},
  034004 (2012)

\bibitem{Karliner2016}
M.~Karliner, S.~Nussinov, J.L. Rosner, Phys. Rev. D \textbf{95}, 034011 (2017)

\bibitem{Wang2017}
Z.G. Wang, Eur. Phys. J. C \textbf{77}, 432 (2017)

\bibitem{Liu2019}
M.S. Liu, Q.F. L\"u, X.H. Zhong, Q.~Zhao, Phys. Rev. D \textbf{100}, 016006
  (2019)

\bibitem{Weng2020}
X.Z. Weng, X.L. Chen, W.Z. Deng, S.L. Zhu, Phys. Rev. D \textbf{103}, 034001
  (2021)

\bibitem{Lundhammar2020}
P.~Lundhammar, T.~Ohlsson, Phys. Rev. D \textbf{102}, 054018 (2020)

\bibitem{Wan2020}
B.D. Wan, C.F. Qiao, Phys. Lett. B \textbf{817}, 136339 (2021)

\bibitem{Zhu2020}
R.~Zhu, Nucl. Phys. B \textbf{966}, 115393 (2021)

\bibitem{Liang2021}
Z.R. Liang, X.Y. Wu, D.L. Yao, Phys. Rev. D \textbf{104}, 034034 (2021)

\bibitem{Wang:2020wrp}
J.Z. Wang, D.Y. Chen, X.~Liu, T.~Matsuki, Phys. Rev. D \textbf{103}, 071503
  (2021)

\bibitem{Wang:2022jmb}
J.Z. Wang, X.~Liu, Phys. Rev. D \textbf{106}, 054015 (2022)

\bibitem{Dong:2020nwy}
X.K. Dong, V.~Baru, F.K. Guo, C.~Hanhart, A.~Nefediev, Phys. Rev. Lett.
  \textbf{126}, 132001 (2021), [Erratum: Phys.Rev.Lett. 127, 119901 (2021)]

\bibitem{Chen2022}
H.X. Chen, W.~Chen, X.~Liu, Y.R. Liu, S.L. Zhu (2022), \texttt{2204.02649}

\bibitem{Biloshytskyi:2022dmo}
V.~Biloshytskyi, V.~Pascalutsa, L.~Harland-Lang, B.~Malaescu, K.~Schmieden,
  M.~Schott (2022), {accepted in Phys. Rev. D}, \texttt{2207.13623}

\bibitem{HarlandLang2019}
L.A. Harland-Lang, V.A. Khoze, M.G. Ryskin, Eur. Phys. J. C \textbf{79} (2019)

\bibitem{HarlandLang2016}
L.A. Harland-Lang, V.A. Khoze, M.G. Ryskin, Eur. Phys. J. C \textbf{76} (2016)

\bibitem{Wang2018}
J.Z. Wang, Z.F. Sun, X.~Liu, T.~Matsuki, Eur. Phys. J. C \textbf{78}, 1 (2018)

\bibitem{Bern:2001dg}
Z.~Bern, A.~De~Freitas, L.J. Dixon, A.~Ghinculov, H.L. Wong, JHEP \textbf{11},
  031 (2001)

\bibitem{Klusek-Gawenda:2016nuo}
M.~K\l{}usek-Gawenda, W.~Sch\"afer, A.~Szczurek, Phys. Lett. B \textbf{761},
  399 (2016)

\bibitem{Krintiras:2022jxa}
G.K. Krintiras, I.~Grabowska-Bold, M.~K\l{}usek-Gawenda, E.~Chapon,
  R.~Chudasama, R.~Granier~de Cassagnac (2022), \texttt{2204.02845}

\bibitem{LHCb2020}
{LHCb collaboration}, Science Bulletin \textbf{65}, 1983 (2020)

\bibitem{Barger1975}
V.~Barger, R.~Phillips, Phys. Lett. B \textbf{58}, 433 (1975)

\bibitem{Redlich2000}
K.~Redlich, H.~Satz, G.M. Zinovjev, Eur. Phys. J. C \textbf{17}, 461 (2000)

\bibitem{Esposito2021}
A.~Esposito, C.A. Manzari, A.~Pilloni, A.D. Polosa, Phys. Rev. D \textbf{104},
  114029 (2021)

\bibitem{Debastiani2019}
V.R. Debastiani, F.S. Navarra, Chin. Phys. C \textbf{43}, 013105 (2019)

\end{thebibliography}

\end{document}